# Thermal conductivity of armchair phosphorene nanotubes: a molecular dynamics study

Feng Hao, Xiangbiao Liao, Hang Xiao, and Xi Chen[*]

Department of Earth and Environmental Engineering, Columbia University, New York, NY 10027, USA

[*]Corresponding author: xichen@columbia.edu

Tel.: +1-212-854-3787

**Abstract**

The effects of size, strain, and vacancies on thermal properties of armchair phosphorene nanotubes are investigated through molecular dynamics simulations. It is found that the thermal conductivity has a remarkable size effect because of the restricted paths for phonon transport, strongly depending on the diameter and length of nanotube. Owing to the intensified low-frequency phonons, axial tensile strain can facilitate thermal transport. On the contrary, compressive strain weakens thermal transport due to the enhanced phonon scattering around the buckling of nanotube. In addition, the thermal conductivity is dramatically reduced by single vacancies, especially upon high defect concentrations.

# 1. Introduction

Phosphorene, a novel two-dimensional (2D) material, has attracted tremendous research interest since it was successfully exfoliated from black phosphorus [1-3]. In contrast to graphene, monolayer phosphorene has a direct bandgap of 1.51 eV [4]. In experiments, it has been observed that multi-layer phosphorene field effect transistors exhibit large current on-off ratios, high charge mobilities, and fast photo-response, opening up more opportunities in electronic devices [3,5,6]. Meanwhile, phosphorene nanoribbon has been considered as a promising candidate for thermoelectric applications due to the enhanced Seebeck coefficient and decreased thermal conductivity compared with its pristine counterpart of 2D phosphorene [7].

Monolayer phosphorene features a puckered structure with each phosphorus atom connected by three adjacent atoms through covalent bonds. Phosphorene has two distinct in-plane directions, i.e. the armchair direction and zigzag direction. Thus, anisotropy is extremely notable regarding its physical properties, such as the Young's modulus [8], thermal conduction [9], and electrical conductance [10].

In practice, thermal transport is of great importance for the electronic and thermoelectric devices. Currently, most studies have been focusing on the thermal conductivity of pristine phosphorene. By using first-principles calculations, it was observed that the thermal conductivity has a significant crystallographic orientation dependence [11-13]. Alternatively, Hong *et al.* and Xu *et al.* separately presented a systematical study of thermal conductivity for 2D phosphorene through molecular dynamics (MD) simulations [14,15]. However, the investigations on phosphorene nanotubes (PNT) are rare, which is the main motivation of this work. It is evident that the thermal conductivity of PNT replies on the sizes of diameter and length. In addition, the effects of strain and point defects on thermal conductivity are explicitly investigated here in view of the following facts.

The electronic properties can be artificially modified by external strains, and this tunability is crucial in nanoelectronics. For example, Rodin *et al.* found that strain can largely change the gap size of phosphorene and induce a semiconductor-metal transition [16], and Guo *et al.* found that armchair PNT undergoes a direct-to-indirection band gap transition under a critical tensile strain [17], and Yu *et al.* found that the hole mobility of armchair PNT greatly increases as the compressive strain is applied [18]. Moreover, the optoelectronic properties strongly depend on strain and local geometry [19]. As strain plays key roles in modulating physical properties of phosphorene, the effects of strain on thermal properties need to be clarified.

Defects are commonly observed in 2D materials, such as graphene and silicene [20,21]. In the experimental techniques of micromechanical cleavage and liquid-phase exfoliation [22,23], point defects potentially exist in the phosphorene flakes. To this end, the defects in phosphorene were theoretically studied based on density functional theory, which confirmed the stability of a wide variety of point defects, such as the Stone-Wales defect, and single and double vacancy defects [24,25]. In addition, phosphorene has a high chemical reactivity and thus is prone to be oxidized when exposed to air, resulting in degradation of the structure and electronic properties [26-29]. In essence, defects alter the atomistic structure of pristine PNT and therefore its thermal properties.

In this work, we perform MD simulations to study the underlying mechanism with regard to the effects of size, strain, and defects on thermal properties, shedding light on the specified thermal applications of PNTs. It should be noted that the total thermal conductivity of phosphorene is dominated by the phonon contribution, and hence electronic contribution is neglected in this study. To begin with, it is emphasized that the scope of this study is limited to the single vacancy (SV) for the impact of defects on thermal conductivity, and we only focus on the armchair phosphorene nanotube (a-PNT) due to its more stable and flexible structure over the zigzag phosphorene nanotube (z-PNT) addressed in the next section.

## 2. Model and Method

The structure of 2D phosphorene is depicted in Fig. 1(a), and the lattice constants of unit cell are $a_1$ = 3.2 Å and $a_2$ = 4.2 Å in the zigzag and armchair directions, respectively. Rolling up the 2D phosphorene along the vector **R** = $n_1\boldsymbol{a}_1+n_2\boldsymbol{a}_2$, where $n_1$ and $n_2$ are the numbers of unit cell along the two directions, it produces the ($n_1,n_2$) PNT. Specifically, Fig. 1(b) shows the (0,10) a-PNT with the zigzag-oriented axial direction, and Fig. 1(c) illustrates the (10,0) z-PNT with the armchair-oriented axis. Nevertheless, a large strain energy is stored in the curved PNT, and the tensile strain of the outer atomic layer is roughly given as

$$\varepsilon_z = \frac{\pi t}{a_1 n_1}, \quad \varepsilon_a = \frac{\pi t}{a_2 n_2} \tag{1}$$

Where $\varepsilon_z$ is for the z-PNT, $\varepsilon_a$ is for the a-PNT, and $t$ is the distance between the two atomic layers shown in Fig. 1(b), which is assumed to be 2.0 Å here. Fig. 4(d) shows that strain greatly increases with decreasing $n$, implying the relative instability of PNT under small diameters. Taking the failure strains of 0.06 (zigzag) and 0.13 (armchair) at temperature $T = 300$ K [30], the estimated critical $n$ are 34 and 12 for the z-PNT and a-PNT, respectively, below which the structure may fracture because of the high strain that the PNT experiences. The crude analysis thus demonstrates that a-PNT is more robust compared to z-PNT. As a result, we only aim at probing the thermal conductivity of a-PNT in the following.

In our studies, all results are obtained through molecular dynamics simulations, as implemented in the LAMMPS package [31]. The Stillinger-Weber (SW) potential is used for the interatomic interactions between phosphorus atoms, predicting an accurate phonon spectrum and mechanical behaviors [30], which is quite sophisticated for this study. To eliminate the coupling effect of neighboring PNTs, a vacuum layer of 100 Å is considered around the a-PNT, while periodic boundary condition is applied along the axial direction.

The structure is firstly equilibrated at ambient condition (temperature $T = 300$ K and pressure $P = 1$ atm) under a Nosé-Hoover thermostat for 500 *ps*, and the time step is 0.5 *fs*. In the non-equilibrium Müller-Plathe approach [32], a-PNT is divided into $2N$ ($N = 10$) layers in the axial direction. The kinetic energies of the hottest atom in layer 1 and the coldest atom in the middle layer $N+1$ are exchanged continually for 5 *ns* to produce a symmetric temperature gradient, thereby forming the equal thermal flux $J$ at both sides separated by the two layers in Fig. 2(a). By satisfying the Fourier's law of heat conduction, the thermal conductivity is thus calculated as

$$\kappa = -J(A\partial T / \partial x)^{-1} \tag{2}$$

where $A$ is the cross-sectional area of a-PNT, which is given as $hna_2$. As shown in Fig. 1(b), the interlayer spacing 5.24 Å of black phosphorus is taken as the a-PNT thickness $h$. $\partial T / \partial x$ is the thermal gradient, which is obtained by linearly fitting the temperature in the linear region shown in Fig. 2(a).

In the tensile or compressive simulations, uniform strain is applied to the equilibrium structure in the nanotube axial direction, with the strain rate of 0.0001/*ps*. Subsequently, the structure is relaxed for 500 *ps* in the NVT ensemble before the Müller-Plathe simulation is carried out to obtain the thermal conductivity of strained a-PNT. The strain investigated ranges from -0.03 to 0.03, where negative strain indicates compression and positive strain represents tension. To assess the effect of defects, monatomic vacancies are randomly placed in the a-PNT, and the defect concentration *f* is defined as the number density of atoms removed from the pristine a-PNT.

## 3. Results and discussion

First of all, the size-dependent thermal conductivity is explored for the optimized a-PNTs. In Fig. 2(b), it shows the thermal conductivity $\kappa$ with various $n$, associated with the diameter of nanotube by $D = na_2/\pi$, at temperature $T = 300$ K. The length of nanotube is chosen to be 100 nm, while periodic boundary condition is applied along the axial direction for each a-PNT. Fig. 2(b) clearly demonstrates that as $n$ increases, $\kappa$ sharply increases and then comes to a gentle rise after $n = 20$. As the characteristic size scales down to the order of several nm, the phonon confinement leads to reduced phonon group velocities and phonon mean-free path (MFP), thereby weakening the thermal transport. In previous studies, this size effect was also observed in silicon nanowires and carbon nanotubes [33,34].

Besides, the thermal conductivity of nanotube strongly depends on the length. To investigate this effect, Fig. 3(a) illustrates the relationship of thermal conductivity and length for the (0,20) a-PNT. As shown in Fig. 3(a), $\kappa$ gradually increases with increasing length. Therefore, the size effect of thermal transport is related to not only the diameter but also the length. As studied here, the length of the simulation system $L$ is not significantly longer than the phonon MFP, resulting in phonon scattering occurring at the interfaces between the heat source and sink. In other words, $\kappa$ is largely limited by the size, called as the Casimir limit. The thermal conductivity of infinite PNT can be figured out by using linear extrapolation, which takes the following form [35]

$$\frac{1}{\kappa} = C(\frac{1}{l_\infty} + \frac{4}{L}) \tag{3}$$

where $l_\infty$ is the phonon MFP of infinite (0,20) a-PNT, and $C$ is a constant. Given that the representative MFP along the zigzag direction is 66 nm for the phosphorene sheet [11], the lengths of 150 - 300 nm are employed for the linear fitting between $1/\kappa$ and $1/L$, since Eq. (1) is valid when $L$ is comparable or larger than the phonon MFP [36]. The fitted relationship reads $1/\kappa = 3.1/L+0.02233$, and hence it yields the estimated thermal conductivity of 44.8 W/mK in the limit $1/L = 0$ and the phonon MFP of infinite (0,20) a-PNT $l_\infty = 35$ nm. In a previous study, Hong *et al*. used the same SW potential and reported the thermal conductivity of 110.7 W/mK along the zigzag direction for phosphorene sheet [14]. Hence, the thermal conductivity of PNT is greatly reduced compared to that of 2D phosphorene, largely originating from the restricted paths for phonon transport in PNTs, as the phonon MFP shrinks down to 35 nm of a-PNT against 66 nm of 2D phosphorene.

It has been demonstrated that strain plays an important role in modulating electronic properties of a-PNT [17,18]. Therefore, the effects of strain on thermal properties are essential to be elucidated. The strain is applied along the nanotube axial direction, ranging from -0.03 to 0.03. In Fig. 4(a), it can be seen that strain has a non-negligible impact on the thermal conductivity. As the a-PNT undergoes compression, the thermal transport is hindered, and the inset depicts the configuration under the compressive strain of -0.02. As a matter of fact, strain induced reduction in thermal conductivity is widely found for flexible structures at the nanoscale, including graphene and carbon nanotube [37]. For low-dimensional materials, compressive force can easily break the structure stability due to the low stiffness of cross-section, producing a large deformation or buckling along the force direction shown in the inset. As a result, phonon scattering is enhanced in the vicinity of buckling. It is worth noting that a-PNT could experience a substantial damage when stress exceeds the ideal strength under the buckling, since the axial direction possesses a relative low failure strain, which is quite consistent with the wrinkling and rupture attributed to the pressure along the zigzag direction [38].

In contrast to the effect of compressive strain, the thermal conductivity of a-PNT monotonically increases with increasing tensile strain. To our intuition, phonon should be softened when the crystalline structure stand a stretching strain, leading to the degradation of thermal transport. More specifically, the group velocities are significantly reduced by tensile strain with respect to silicon [37]. As for graphene, however, the mechanism of $\kappa$ reduction is that the out-of-plane phonon mode is greatly suppressed due to tension [37]. Similarly, tensile strain can also induce considerable changes in phonon dispersion of 2D phosphorene, particularly for the acoustic modes LA, ZA, and LA [39], which occupy the major contribution to thermal conductivity. However, on the contrary, stretching the a-PNT along the zigzag direction facilitates phonon transport. By using first-principles calculations and the nonequilibrium Green's function method, Ong *et al*. found the similar results that the zigzag-oriented thermal conductance is enhanced by a zigzag-oriented tensile strain for 2D phosphorene, and revealed that the change of phonon transmission accounts for this anomalous effect [12].

To further unveil the mechanism, phonon density of state (PDOS) is employed, which is determined by the Fourier transform of the velocity $v$ autocorrelation function

$$F(\omega) = \frac{1}{\sqrt{2\pi}} \int_{-\infty}^{\infty} \frac{<v(t)v(0)>}{<v(0)v(0)>} e^{i\omega t} dt \tag{4}$$

In Fig. 4(b), the PDOS of (0,20) a-PNT is truncated at the frequency of 260 cm$^{-1}$, since the contribution by the high-frequency optical phonons to thermal conductivity can be neglected [12]. It is demonstrated that the low-frequency phonons ($\omega < 70$ cm$^{-1}$) are prominently intensified under tensile strain compared to the free-standing a-PNT, especially for the zigzag direction along which heat flux transports. That is, long-wave phonons can be activated as the a-PNT is stretched along the axial direction, leading to the increase in thermal conductivity.

Finally, the influence of point defects on thermal conductivity is explored. Defects inevitably exist within the phosphorene, such as oxygen chemisorption and vacancies [24-29]. Here, we focus on the effect of single vacancy (SV) on thermal properties. The defects are randomly distributed in the (0,20) a-PNT with the length of 60 nm, and the defect concentration $f$ is up to 1%. To ensure the thermal stability of defected PNTs, the thermal conductivity is calculated at temperature $T = 250$ K. As shown in Fig. 4(a), $\kappa$ is greatly sensitive to defects, indicating a striking decrease. For instance, $\kappa$ is reduced by 61% under $f = 1\%$. The localized SV defects serve as phonon scattering centers to heat flux through them, strongly lowering the phonon MFP, which further induces the drastic reduction in thermal conductivity. This effect has also been observed for graphene and carbon nanotube [40,41]. In comparison with graphene [40], the thermal conductivity of a-PNT has a less aggressive reduction in the presence of SVs. This is because the phonon MFP of pristine phosphorene is intrinsically shorter than that of graphene [14], and defects thus produce a weaker phonon scattering with respect to phosphorene.

In Fig. 5(b), the temperature dependence of thermal conductivity is illustrated, including the defected (0,20) a-PNT with $f = 0.5\%$ and its counterpart of pristine a-PNT. Obviously, $\kappa$ decreases with temperature from 100 K to 300 K, stemming from more intense phonon-phonon scatterings at high temperatures. According to the $T^{-\alpha}$ law, Fig. 5(b) suggests that the power factors are $\alpha = 0.64$ and $\alpha = 0.60$ for the defect-free a-PNT and defected one, respectively. The trend of temperature-dependent $\kappa$ here is in good agreement with the recent experiments, which measured the in-plane thermal conductivity of black phosphorus nanoribbon [42]. Nevertheless, MD simulations fail to capture the declining trend of thermal conductivity in low-temperature regime $T < 100$ K.

Apart from the potential in electronic devices, phosphorene has recently been proposed as a candidate for thermoelectric material. As studied here, the phonon mean free path could be decreased because of the strong phonon scattering by defects, which leads to the reduction in thermal conductivity. It can be imagined that the influence of defects is not limited to a-PNT, and it takes effect on 2D phosphorene. Moreover, considering that phosphorene is subjected to strain under working conditions, such as thermal mismatch with substrate and external forces, the impact of strain on thermal properties should be taken into account. Thus, this work provides a fundamental study of a-PNT to understand its thermal properties at nanoscale, and is expected to provide a guideline for thermal issues in device design. However, chemical reactions take place when phosphorene is exposed to air [26], which needs to be further considered regarding the thermal transport of phosphorene.

## 4. Conclusion

In summary, we investigate the thermal properties of phosphorene nanotubes by performing molecular dynamics simulations, with emphasis on the effects of size, strain, and single vacancy defect. Compared to the 2D phosphorene, the thermal conductivity of a-PNT evidently decreases due to phonon confinement, exhibiting remarkable size effects, which relies on not only the diameter but only the length. In practice, phosphorene experiences deformation under certain circumstances. Our results indicate that the thermal conductivity of a-PNT decreases under compressive strain because of the buckling in the axial direction. Conversely, owing to the applied stretching strain, the intensified low-frequency phonons contribute to the increase in thermal conductivity. In addition, defects have critical impacts on thermal performance due to the strong phonon scattering around defects, which should affect phosphorene as a potential thermoelectric material.

**FIGURES AND FIGURE CAPTIONS**

**Fig. 1**. Schematic construction of PNTs from 2D phosphorene. Following the roll-up vector **R** = $n_1\boldsymbol{a}_1+n_2\boldsymbol{a}_2$, it can form the (0,10) a-PNT and (10,0) z-PNT in (b) and (c), repectively. (d) A rough prediction of the critical *n*, above which PNTs have robust structures.

**Fig. 2**. (a) In the Muller-Plathe method, the a-PNT is divided into 2*N* layers, with the cold layer 1 and hot layer *N*+1. (b) The diameter-dependent thermal conductivity of a-PNT with a specific length of 100 nm.

**Fig. 3**. The thermal conductivity *κ* of a-PNT with varying axial length under temperature *T* = 300 K. (b) Linear fitting of the relationship between the inverse of *κ* and the inverse of *L*.

**Fig. 4**. (a) The effect of axial strain on the thermal conductivity of a-PNT, and the buckled nanotube under a compressive strain of -0.02 is insert. (b) The comparison of phonon density of state (PDOS) between the free-standing a-PNT and the one with a tensile strain of 0.03.

**Fig. 5**. (a) Thermal conductivity of a-PNT with increasing single vacancy defect concentration at temperature *T* = 250 K. In (b), the comparison of temperature-dependent thermal conductivity for the pristine a-PNT and defected a-PNT with the defect concentration of 0.5%.

Figure 1

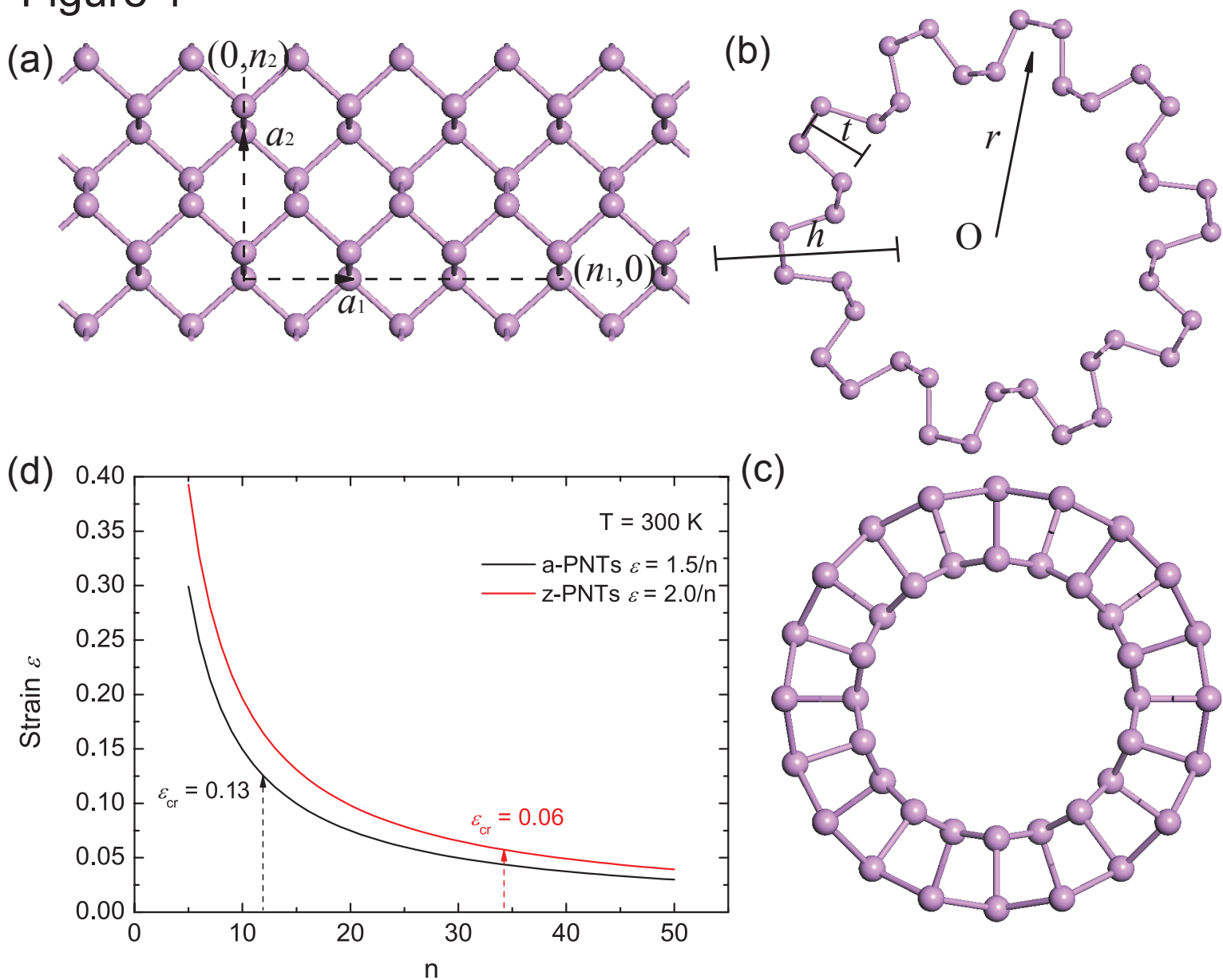

Figure 2

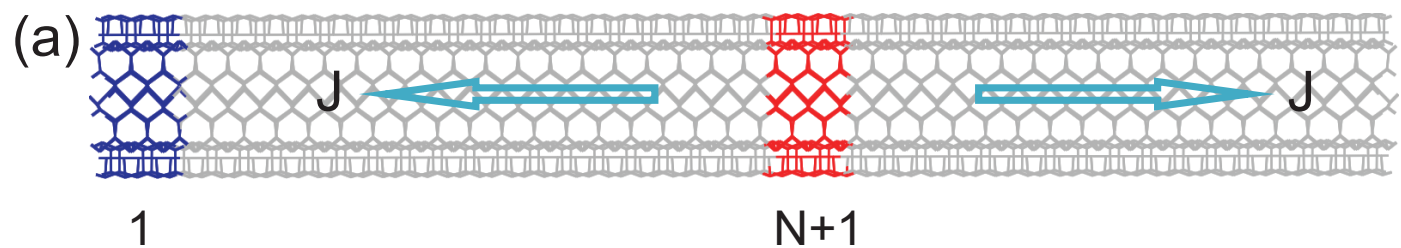

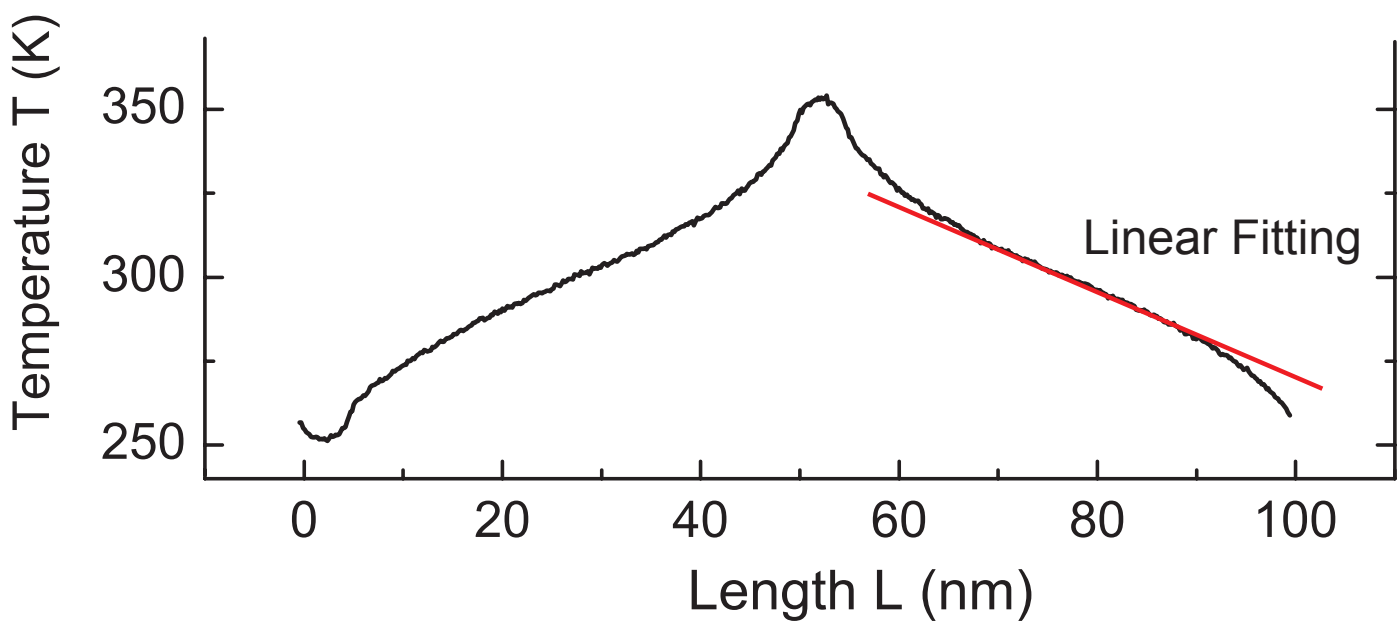

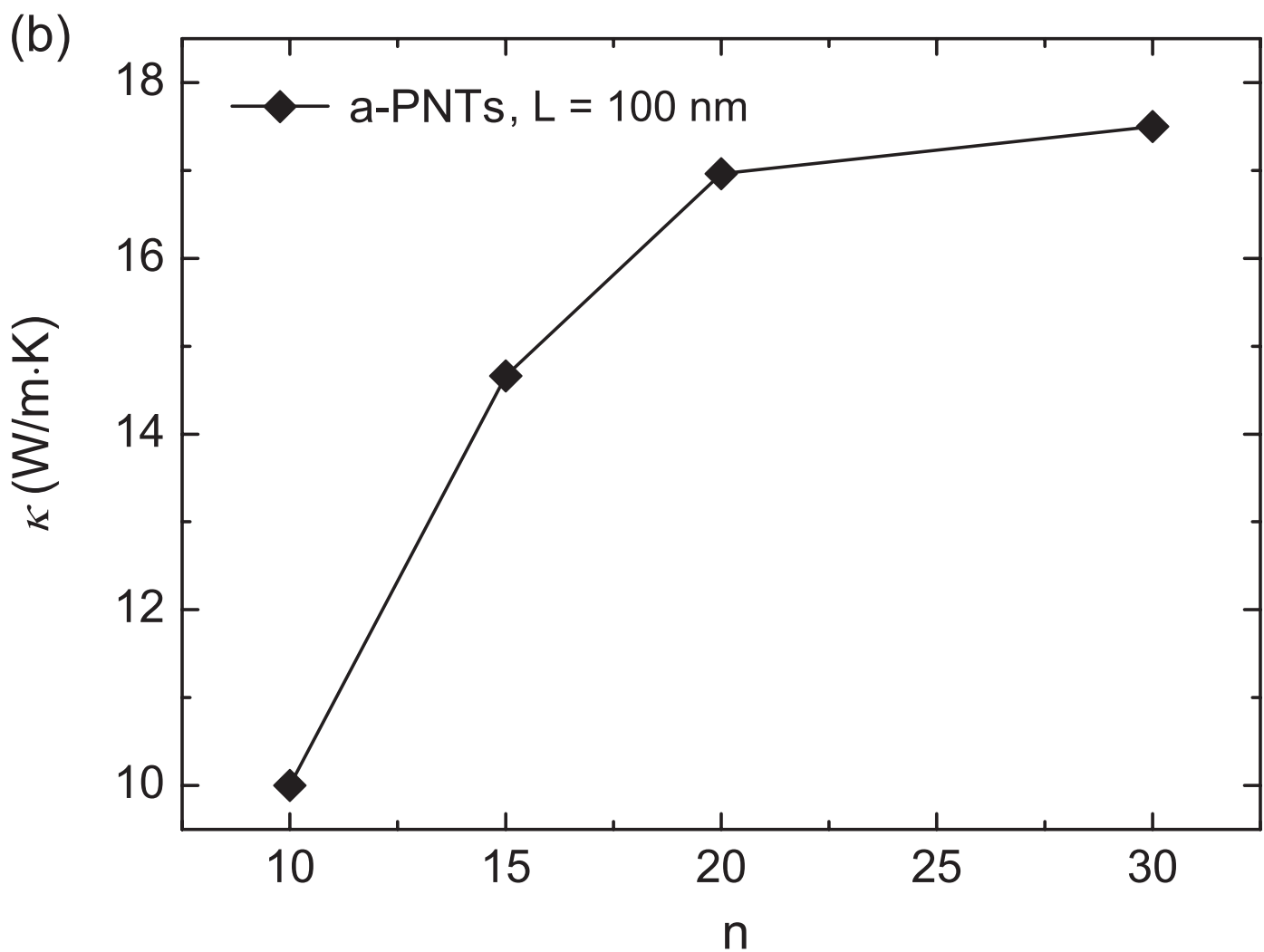

Figure 3

(a)
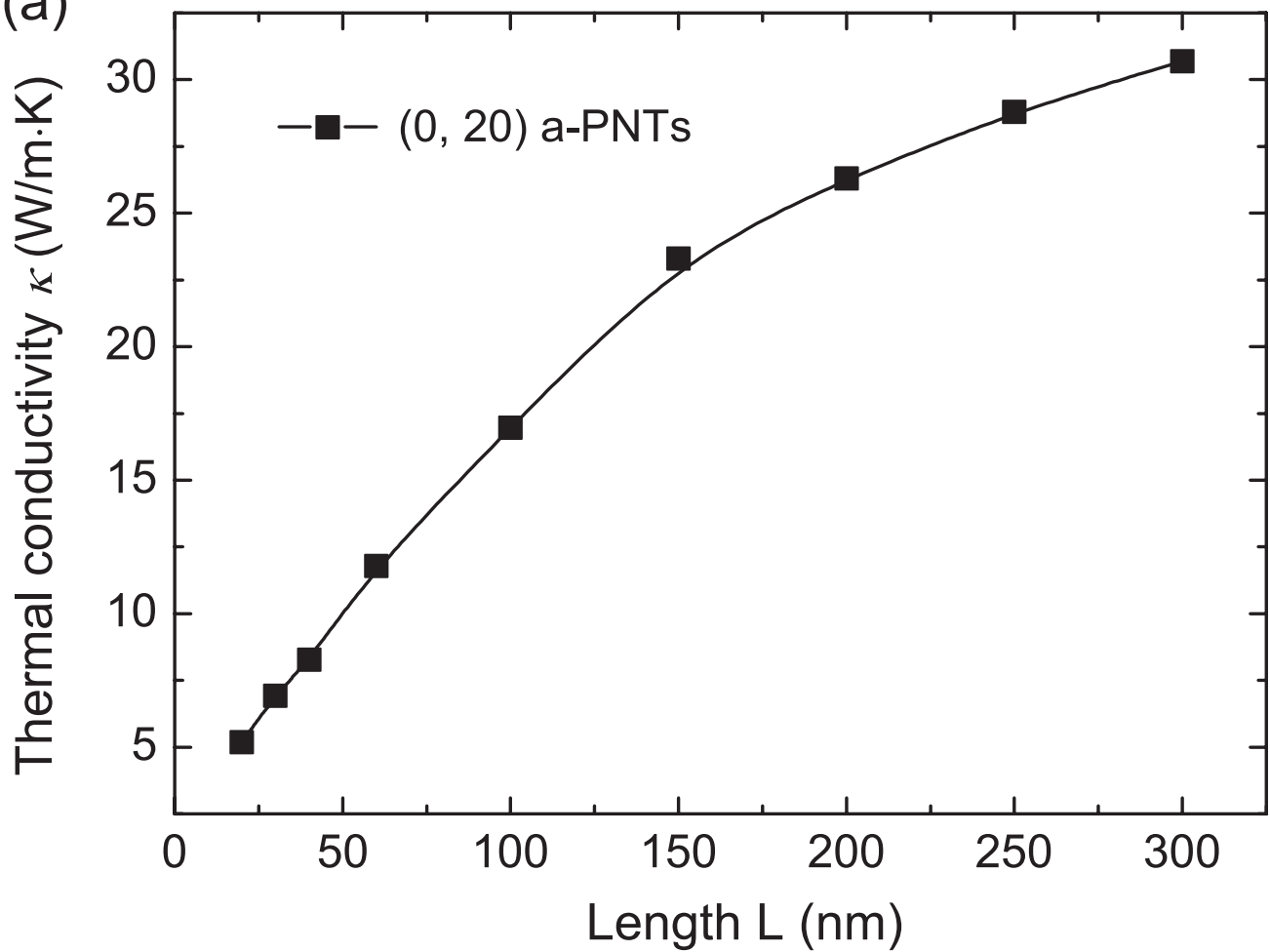

(b)
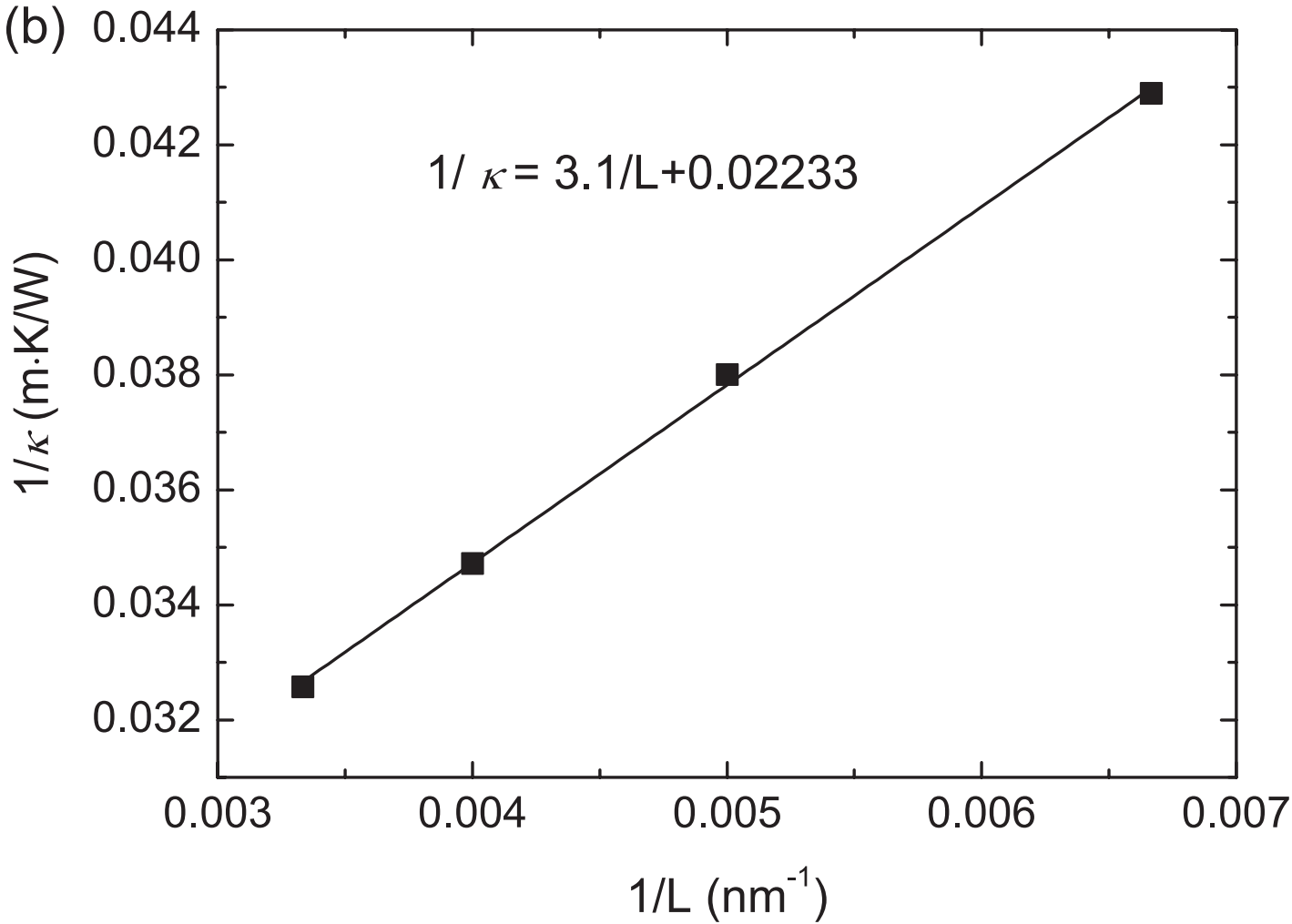

$1/\kappa = 3.1/L + 0.02233$

Figure 4

(a)

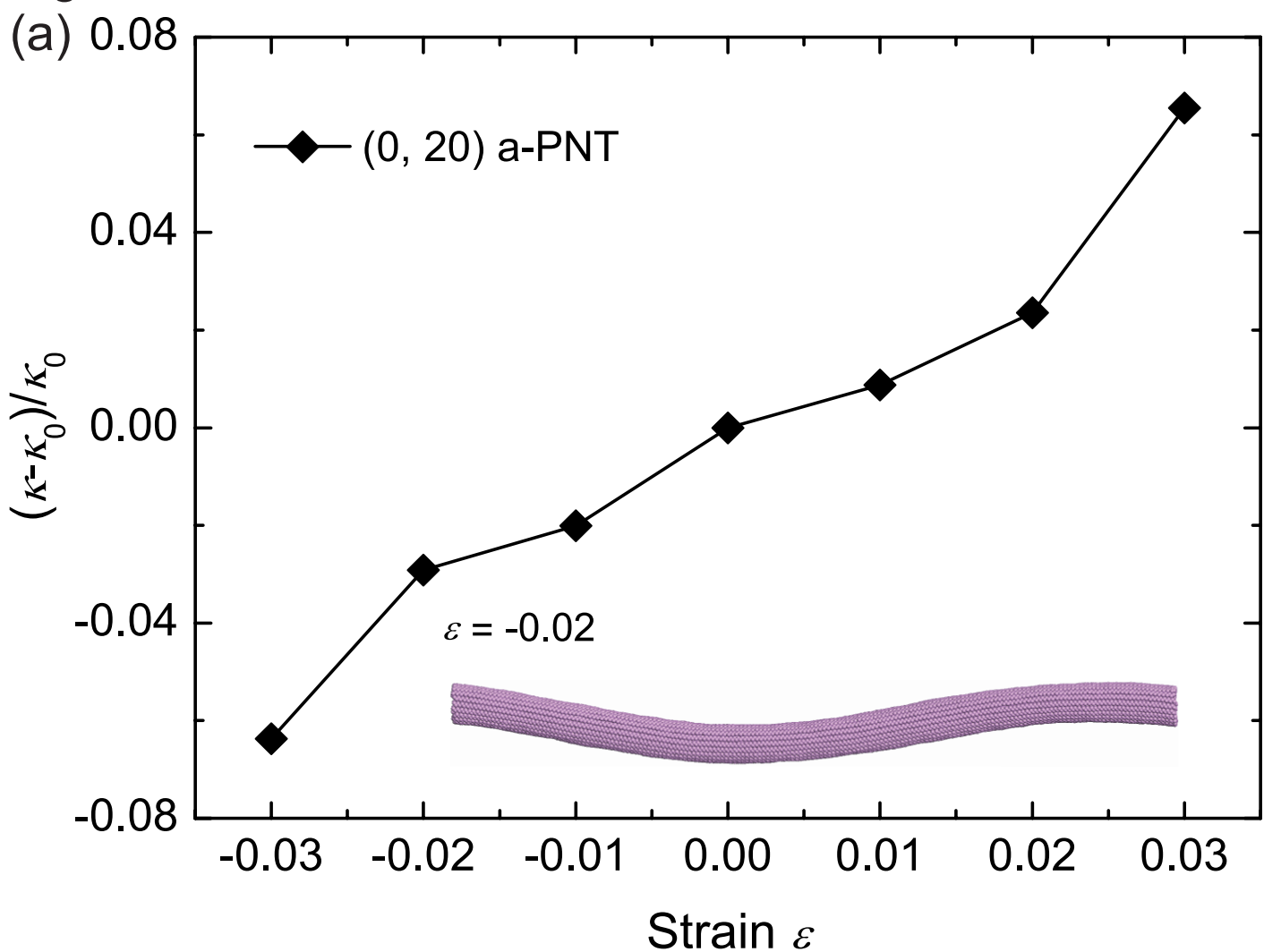

(b)

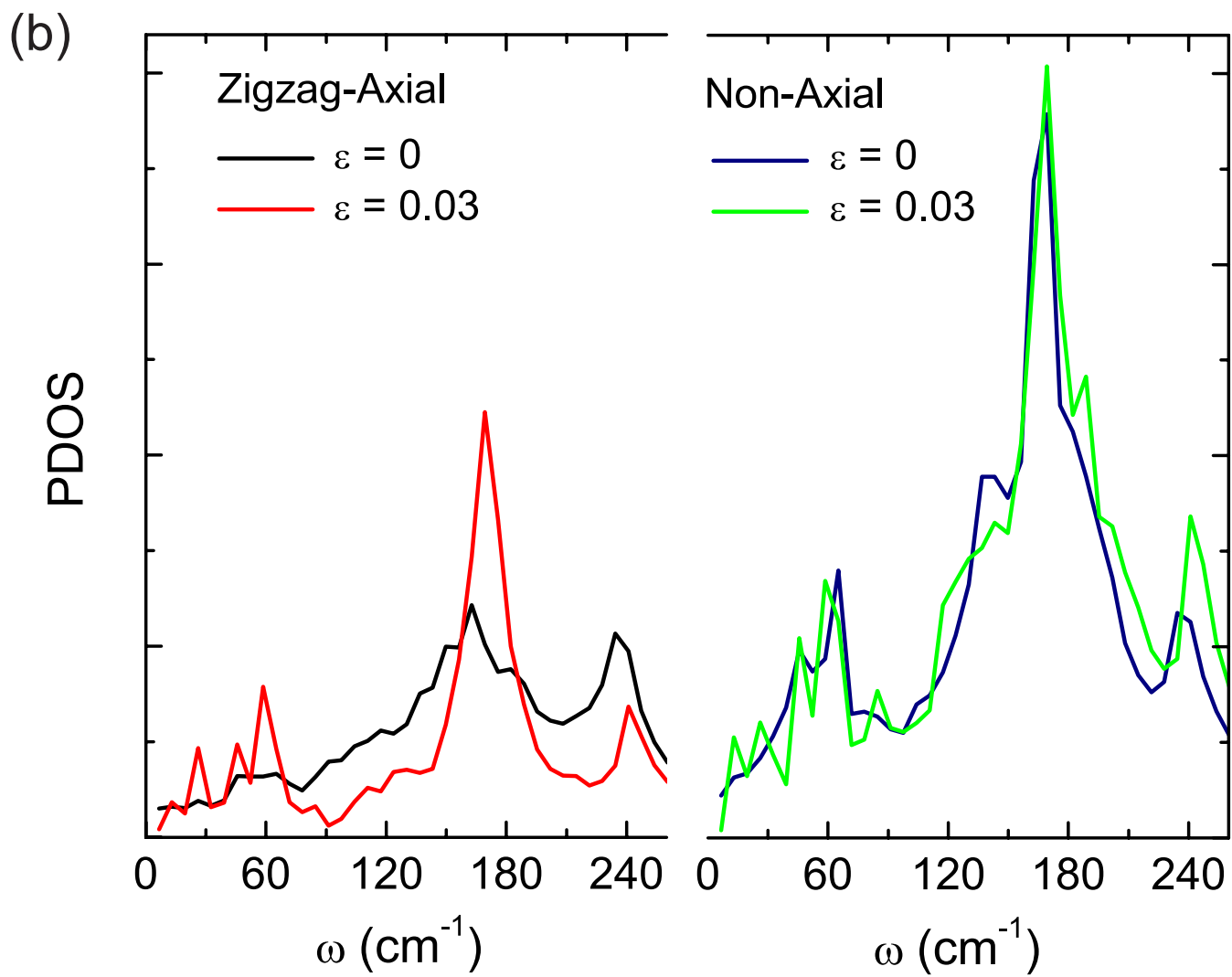

Figure 5

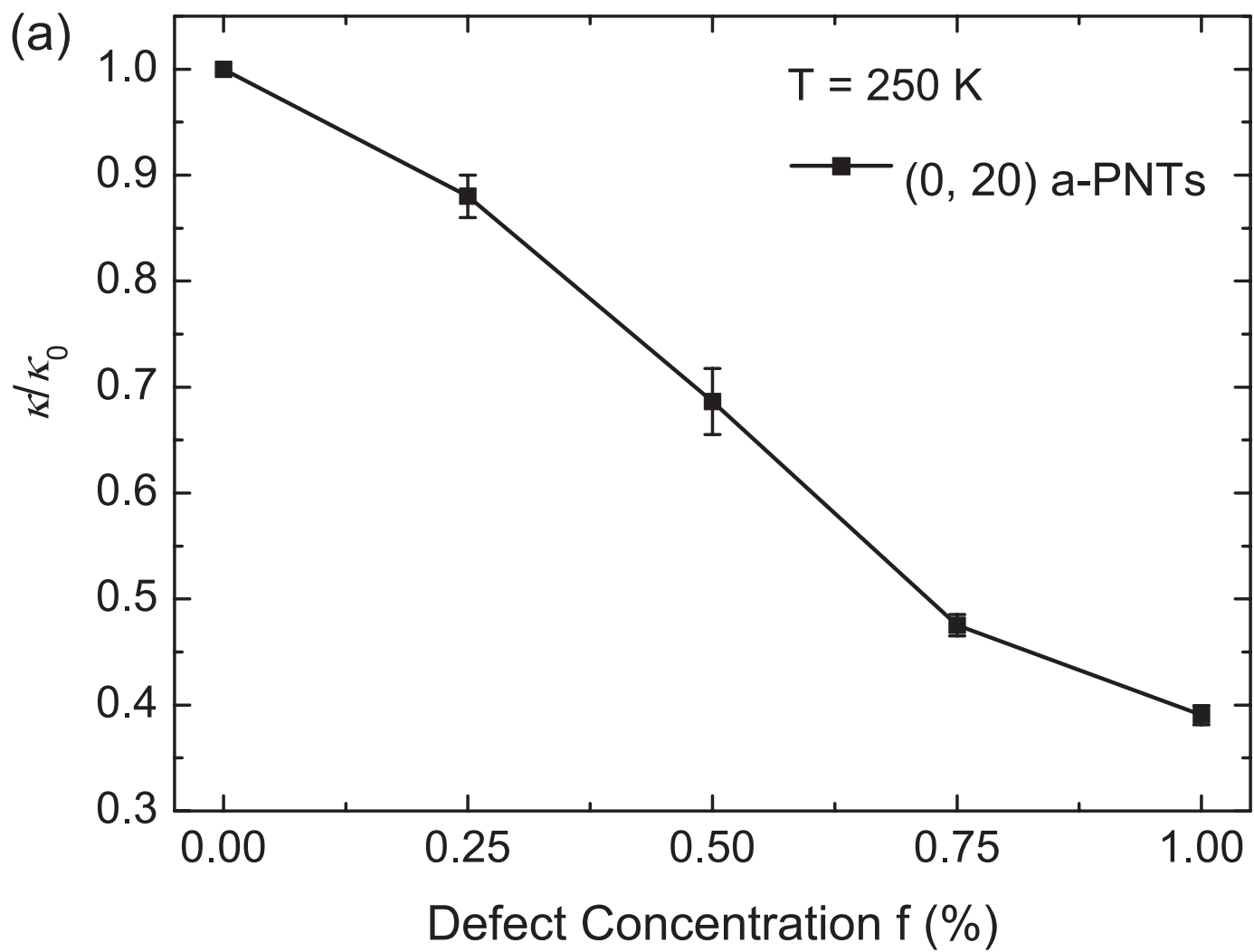

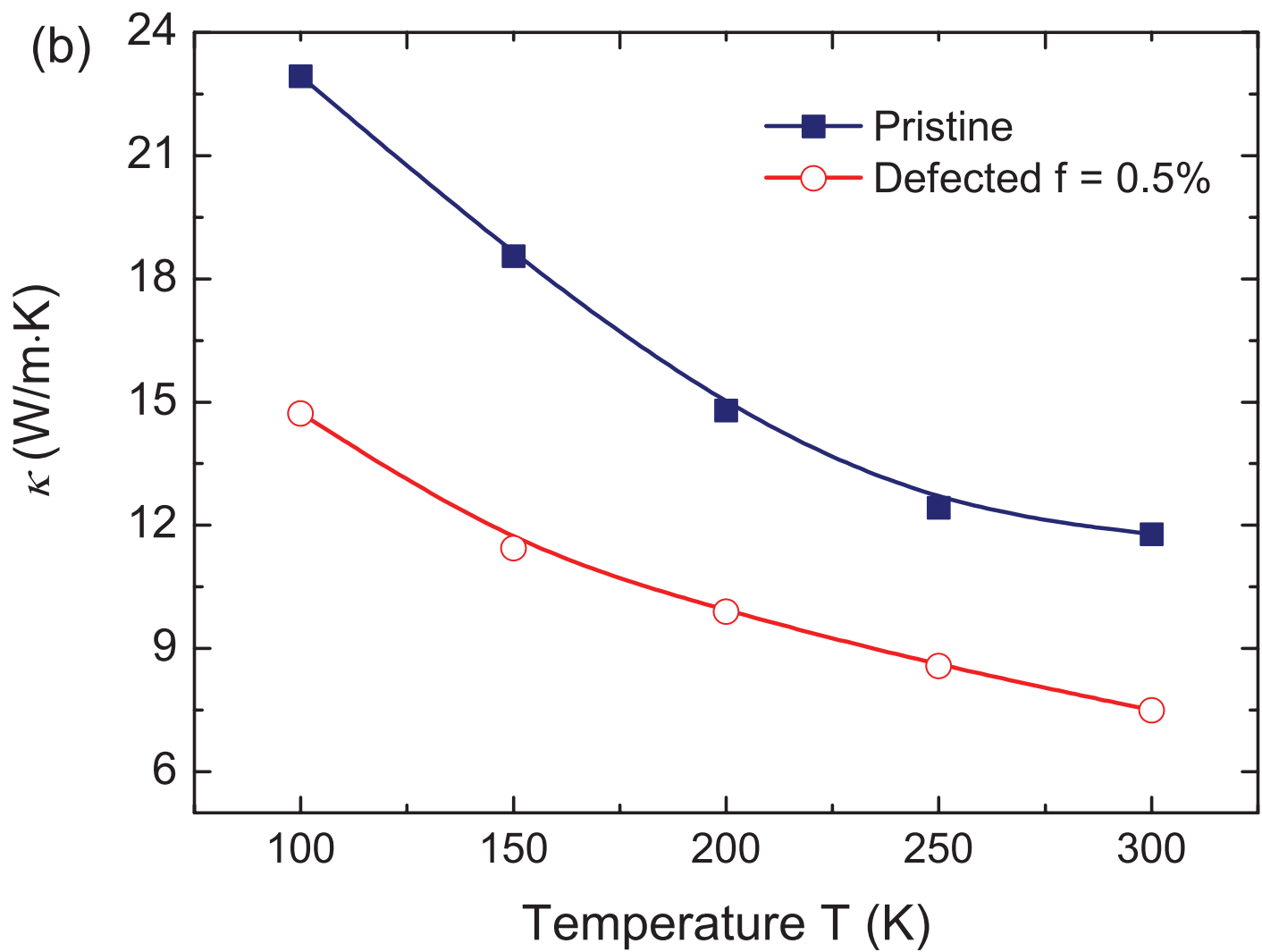